\documentclass[letterpaper,twocolumn,prl,aps,superscriptaddress,amsmath,amssymb,floatfix]{revtex4-1}
\usepackage{mathptmx}
\DeclareMathAlphabet{\mathcal}{OMS}{cmsy}{m}{n}
\usepackage[latin9]{inputenc}
\usepackage{color}
\usepackage{verbatim}
\usepackage{float}
\usepackage{amsmath}
\usepackage{amssymb}
\usepackage{graphicx}
\usepackage{esint}
\usepackage[unicode=true,
 bookmarks=true,bookmarksnumbered=false,bookmarksopen=false,
 breaklinks=false,pdfborder={0 0 1},backref=false,colorlinks=true]
 {hyperref}
\hypersetup{
 linkcolor=magenta,urlcolor=blue,citecolor=blue,pdfstartview={FitH},hyperfootnotes=false}

\makeatletter

\usepackage{textcomp}
\usepackage{epstopdf}

\pdfpageheight\paperheight
\pdfpagewidth\paperwidth

\@ifundefined{textcolor}{}{%
 \definecolor{BLACK}{gray}{0}
 \definecolor{WHITE}{gray}{1}
 \definecolor{RED}{rgb}{1,0,0}
 \definecolor{GREEN}{rgb}{0,1,0}
 \definecolor{BLUE}{rgb}{0,0,1}
 \definecolor{CYAN}{cmyk}{1,0,0,0}
 \definecolor{MAGENTA}{cmyk}{0,1,0,0}
 \definecolor{YELLOW}{cmyk}{0,0,1,0}
}

\usepackage{xcolor}
\usepackage{soul}
\newcommand{\bra}[1]{\ensuremath{\left\langle#1\right|}}
\newcommand{\ket}[1]{\ensuremath{\left|#1\right\rangle}}

\definecolor{blue}{rgb}{0,0,1}
\definecolor{red}{rgb}{1,0,0}
\definecolor{green}{rgb}{0,1,0}

\@ifundefined{textcolor}{}{%
 \definecolor{BLACK}{gray}{0}
 \definecolor{WHITE}{gray}{1}
 \definecolor{RED}{rgb}{1,0,0}
 \definecolor{GREEN}{rgb}{0,1,0}
 \definecolor{BLUE}{rgb}{0,0,1}
 \definecolor{CYAN}{cmyk}{1,0,0,0}
 \definecolor{MAGENTA}{cmyk}{0,1,0,0}
 \definecolor{YELLOW}{cmyk}{0,0,1,0}
}

\usepackage{xcolor}\usepackage{soul}
\definecolor{blue}{rgb}{0,0,1}
\definecolor{red}{rgb}{1,0,0}
\definecolor{green}{rgb}{0,1,0}

\usepackage{soul}

\makeatother

\begin{document}
\title{Error-detectable Universal Control for High-Gain Bosonic Quantum Error Correction}

\affiliation{CAS Key Laboratory of Quantum Information, University of Science and Technology of China, Hefei 230026, China}

\affiliation{Center for Quantum Information, Institute for Interdisciplinary Information
Sciences, Tsinghua University, Beijing 100084, China}

\affiliation{Hefei National Laboratory, Hefei 230088, China}

\author{Weizhou~Cai}
\thanks{These authors contributed equally to this work.}
\affiliation{CAS Key Laboratory of Quantum Information, University of Science and Technology of China, Hefei 230026, China}

\author{Zi-Jie Chen}
\thanks{These authors contributed equally to this work.}
\affiliation{CAS Key Laboratory of Quantum Information, University of Science and Technology of China, Hefei 230026, China}

\author{Ming Li}

\affiliation{CAS Key Laboratory of Quantum Information, University of Science and Technology of China, Hefei 230026, China}
\affiliation{Hefei National Laboratory, Hefei 230088, China}

\author{Qing-Xuan Jie}

\affiliation{CAS Key Laboratory of Quantum Information, University of Science and Technology of China, Hefei 230026, China}

\author{Xu-Bo Zou}
\affiliation{CAS Key Laboratory of Quantum Information, University of Science and Technology of China, Hefei 230026, China}
\affiliation{Hefei National Laboratory, Hefei 230088, China}

\author{Guang-Can Guo}
\affiliation{CAS Key Laboratory of Quantum Information, University of Science and Technology of China, Hefei 230026, China}
\affiliation{Hefei National Laboratory, Hefei 230088, China}

\author{Luyan~Sun}
\email{luyansun@tsinghua.edu.cn}
\affiliation{Center for Quantum Information, Institute for Interdisciplinary Information
Sciences, Tsinghua University, Beijing 100084, China}
\affiliation{Hefei National Laboratory, Hefei 230088, China}

\author{Chang-Ling~Zou}
\email{clzou321@ustc.edu.cn}
\affiliation{CAS Key Laboratory of Quantum Information, University of Science and Technology of China, Hefei 230026, China}
\affiliation{Hefei National Laboratory, Hefei 230088, China}

\date{\today}

\begin{abstract}
Protecting quantum information through quantum error correction (QEC) is a cornerstone of future fault-tolerant quantum computation. However, current QEC-protected logical qubits have only achieved coherence times about twice those of their best physical constituents. Here, we show that the primary barrier to higher QEC gains is ancilla-induced operational errors rather than intrinsic cavity coherence. To overcome this bottleneck, we introduce error-detectable universal control of bosonic modes, wherein ancilla relaxation events are detected and the corresponding trajectories discarded, thereby suppressing operational errors on logical qubits. For binomial codes, we demonstrate universal gates with fidelities exceeding $99.6\%$ and QEC gains of $8.33\times$ beyond break-even. Our results establish that gains beyond $10\times$ are achievable with state-of-the-art devices, establishing a path toward fault-tolerant bosonic quantum computing.
\end{abstract}


\maketitle

\noindent\textit{Introduction.-} A central goal of fault-tolerant quantum computing is to realize logical qubits whose effective coherence and operation fidelity substantially surpass those of the underlying physical device~\cite{Nielsen,Shor1995,Shor1996FT,DiVincenzoPRL1996,GottesmanPRA1998,preskill1998,KITAEV20032,Gidney2021howtofactorbit,gidney2025factor2048bitrsa}. The fundamental principle of quantum error correction (QEC) exploits Hilbert space redundancy: by encoding quantum information across multiple physical degrees of freedom, errors can be detected and corrected without disturbing the encoded information. However, this redundancy comes at a cost of the effective error rate of logical qubits, and the genuine QEC protection of logical qubit with its coherence time surpasses the lifetime of constitute physical qubits has remained challenge. Recent experiments have demonstrated that QEC can extend the lifetime of an encoded logical qubit beyond that of the best physical qubit, reaching the break-even regime~\cite{Ofek2016,Ni2023Nature,Sivak2023,Google2024}.
Moving from this early break-even regime to practically useful fault tolerance, however, requires substantially larger QEC gains~\cite{chen2023QEC}.

Bosonic modes are among the most foundational physical systems for quantum information processing, offering numerous benefits. Their infinite Hilbert space is advantageous for transferring large amounts of information in quantum communication and providing redundant degrees of freedom for encoding QEC codes~\cite{ChuangPRA1997,BraunsteinPRL1998,Gottesman2001PRA,CochranePRA1999,Michael2016PRX,AlbertPRA2018,cai2020bosonic,MA2021Bosonic,Joshi2021Bosonic}. This approach yielded the first demonstration of break-even QEC using cat codes~\cite{Ofek2016}, subsequently extended to binomial codes~\cite{Ni2023Nature} and Gottesman-Kitaev-Preskill grid states~\cite{Sivak2023}, with the essential universal control of the bosonic logical qubit through an ancillary transmon qubit. Despite these milestones, the achieved QEC gains have been restricted to about twice the break-even point even after a decade since the first demonstration, revealing the fundamental limitations from the imperfect ancilla qubit. The limitation can be understood through a simple error budget analysis. Repetitive QEC protection of a bosonic logical qubit proceeds by periodically applying syndrome detection and correction operations with interval time $t_{\mathrm{int}}$. For the dominant single-photon error of cavity at rate $\kappa$, a first-order QEC code suppresses the logical error to $\alpha_W(\kappa t_{\mathrm{int}})^2$, where $\alpha_W > 1$ reflects enhanced error susceptibility from utilizing a larger Hilbert space. However, each QEC cycle introduces an operational error $\varepsilon_{\mathrm{op}}$ arising from imperfections in the ancilla qubit. The effective logical error rate thus becomes $\varepsilon_{\mathrm{op}}/t_{\mathrm{int}} + \alpha_W \kappa^2 t_{\mathrm{int}}$. Minimizing over $t_{\mathrm{int}}$ yields an optimal error rate of $2\kappa\sqrt{\alpha_W \varepsilon_{\mathrm{op}}}$, giving the maximum achievable QEC gain compared with the physical qubit error rate $\alpha \kappa$ as
\begin{equation}
G_{\mathrm{break}} \sim \frac{\alpha}{2\sqrt{\alpha_W \varepsilon_{\mathrm{op}}}}.
\label{Eq1}
\end{equation}
This scaling reveals the fundamental bottleneck: the QEC gain is limited not by the cavity coherence, but by the ancilla-induced operation errors $\varepsilon_{\mathrm{op}}$. Achieving order-of-magnitude QEC gains therefore demands advanced quantum control strategies that suppress ancilla errors, rather than further improvements in materials alone.

\begin{figure}
\centering{}\includegraphics[width=\columnwidth]{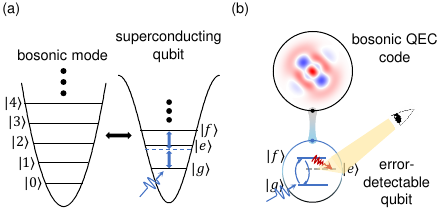}
\caption{Concept of error-detectable (ED) universal control on bosonic modes. (a) A bosonic mode is coupled with an ED ancilla. (b) The ED process of a bosonic mode is realized by error-detecting the ED ancilla and post-selecting the no-error trajectories.}
\label{fig1}
\end{figure}

In this Letter, we introduce error-detectable (ED) universal control of bosonic modes that suppresses dominant ancilla-induced operation errors by converting them into detectable events and post-selecting the no-error trajectories. Numerical simulations demonstrate a universal gate set for binomial codes with process fidelities exceeding 99.5\%, and repeated ED QEC cycles achieving gains of $8.33\times$ beyond break-even under realistic parameters. Finally, we explore dominant errors for ED QEC cycles and reveal a crossover behavior, increasing the ancilla lifetime improves the QEC gain up to a critical regime, beyond which further improvements provide diminishing returns. Our analysis establishes that gains exceeding $10\times$ are achievable with state-of-the-art devices, providing a concrete pathway toward practically useful fault-tolerant quantum computing with bosonic modes.

\smallskip{}
\noindent\textit{ED universal control.-} Figure~\ref{fig1}a sketches the coupled bosonic mode and ancilla transmon qubit system for the hardware-efficient bosonic QEC. By simultaneously driving the bosonic mode and the qubit, universal control of the system can be realized, enabling the necessary encoding, decoding, error syndrome measurement, and recovery operations of the logical qubit~\cite{Krastanov2015,HeeresPRL2015,Heeres2017}. In this work, we focus on the binomial code~\cite{Michael2016PRX,Hu2019,Ni2023Nature}, which have even parity for code space $\mathrm{span}\{\ket{0_L}=(\ket{0}+\ket{4})/\sqrt{2},\, \ket{1_L}=\ket{2}\}$. This code can correct the single-photon-loss error that projects the logical qubit into the odd-parity error space $\mathrm{span}\{\ket{0_E}=\ket{3},\,\ket{1_E}=\ket{1}\}$, with $\ket{n}$ ($n\in\mathbb{Z}$) denoting the Fock state of the mode. In state-of-the-art devices, the ancilla has a significantly shorter lifetime than the bosonic modes by approximately two orders of magnitude~\cite{ReagorPRB2016,Milul2023PRXQ}, which limits the operational fidelity on the bosonic modes. For example, using the numerical optimization via Gradient Ascent Pulse Engineering~\cite{Khaneja2005,Heeres2017} (GRAPE), the achievable QEC operation error $\epsilon_{\mathrm{op}}\sim0.05$, leading to a limited $G_{\mathrm{break}}\sim1$, according to $\alpha_\mathrm{W}=1.27$ for the binomial code and $\alpha=0.6$ for the physical qubit encoding on Fock states \{$\ket{0}$, $\ket{1}$\}.

Our central strategy is to detect the dominant ancilla error, i.e., the longitudinal relaxation, by employing higher energy levels for redundancy. Consequently, the occurrence of ancilla error is detectable and the corresponding experimental trajectories can be discard to mitigate the dominant ancilla error contribution to $\epsilon_{\mathrm{op}}$. We implement this by employing the $\{\ket{g}, \ket{f}\}$ states~\cite{Kubica2023} as the qubit, while using the intermediate state $\ket{e}$ for error detection. The universal control with the ED ancilla can be realized by extending GRAPE to the two-photon drive on the ancilla [Fig.~\ref{fig1}(a)], which generates an effective Hamiltonian~\cite{james2007effective} of the form
\begin{align}
&\mathcal{V}_{\mathrm{A}}/\hbar=\frac{\sqrt{2}(\Omega_{\mathrm{y}}^2-\Omega_{\mathrm{x}}^2)}{2E_{\mathrm{c}}}\sigma_{\mathrm{gf}}^{\mathrm{x}}-\frac{\sqrt{2}\Omega_{\mathrm{x}}\Omega_{\mathrm{y}}}{E_{\mathrm{c}}}\sigma_{\mathrm{gf}}^{\mathrm{y}},
\label{eq:effective_simp_universal}
\end{align}
where $\Omega_{\mathrm{x}},\Omega_{\mathrm{y}}$ denote the drive quadratures along the axes $x,y$ of the ancilla, and $\sigma_{\mathrm{gf}}^{\mathrm{x,y}}$ are the Pauli operators (see Supplementary Materials~\cite{SI} for more details). Then, combined with the dispersive interaction with the bosonic mode, the universal control of the composite system can be realized through the Hamiltonian
\begin{align}
&\mathcal{V}_{\mathrm{B}}/\hbar=-\chi_{\mathrm{e}}a^{\dagger}a\ket{e}\bra{e}-\chi_{\mathrm{f}}a^{\dagger}a\ket{f}\bra{f}+\Omega_{\mathrm{rx}}\frac{a^{\dagger}+a}{2}\nonumber\\
&+\Omega_{\mathrm{ry}}\frac{ia^{\dagger}-ia}{2}+\frac{\sqrt{2}(\Omega_{\mathrm{y}}^2-\Omega_{\mathrm{x}}^2)}{2E_{\mathrm{c}}}\sigma_{\mathrm{gf}}^{\mathrm{x}}-\frac{\sqrt{2}\Omega_{\mathrm{x}}\Omega_{\mathrm{y}}}{E_{\mathrm{c}}}\sigma_{\mathrm{gf}}^{\mathrm{y}}.
\label{eq:H_eff_bosonic}
\end{align}
Here, $\chi_{\mathrm{e,f}}$ denote the dispersive shifts and $\Omega_{\mathrm{rx,ry}}$ are the drive strengths applied to the bosonic mode.

\begin{figure}
\centering{}\includegraphics[width=\columnwidth]{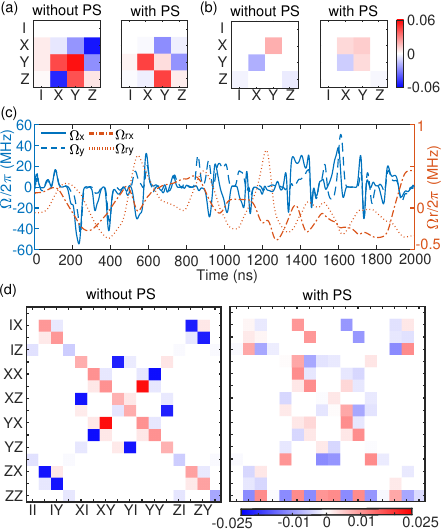}
\caption{Performance of an ED universal gate set of logical qubits. (a-b) Differences between simulated and ideal Pauli transfer matrices for a Hadmard (a) and T (b) gate of a binomial code, respectively. The left and right matrices of each sub figures represent situations without (left) and with (right) error-detection and PS of the ED ancilla. (c) Numerical optimized waveform of a logical Hadmard gate. (d) Differences between simulated and ideal Pauli transfer matrices for a logical controlled-phase gate for two binomial codes with (right) and without (left) error-detection and PS of the ED ancilla.}
\label{fig2}
\end{figure}

The ED ancilla promises the suppression of $\epsilon_{\mathrm{op}}$ by projectively measuring the ancilla in the $\{\ket{g},\ket{e},\ket{f}\}$ basis after operations~\cite{Magnard2018PRL}. To evaluate the performance of the ED universal control on a bosonic mode, we numerically study the gate set on a binomial logical qubit~\cite{Michael2016PRX,Hu2019,Ni2023Nature}, with the experimental parameters: $\chi_{\mathrm{e}}/2\pi=1\,\mathrm{MHz}$, $\chi_{\mathrm{f}}/2\pi=2\,\mathrm{MHz}$, the anharmonicity of the ancilla $E_{\mathrm{c}}/2\pi=400\,\mathrm{MHz}$, the decay rate of the $\ket{e}$ state $\kappa_{\mathrm{e}}=1/40\,\mathrm{\mu s}$, the decay rate of the $\ket{f}$ state $\kappa_{\mathrm{f}}=1/20\,\mathrm{\mu s}$, and the single-photon-loss rate of the bosonic mode $\kappa=1/2\,\mathrm{ms}$. Pure dephasing noise, as described in Refs.~\cite{Scigliuzzo2020,Xu2018}, is not considered in our analysis, as the pure dephasing rate can be only $1\,\mathrm{ms}^{-1}$ for the ancilla qubit~\cite{Ni2023Nature}.


\begin{table}[t]
\centering
\caption{Process fidelities for ED control on binomial logical qubits. PS: post-selection; $P_\mathrm{succ}$: success probability. }
\label{tab:gate_performance}
\begin{ruledtabular}
\begin{tabular}{lccc}
Operation & w/o PS (\%) & w/ PS (\%) & $P_\mathrm{succ}$ \\
\hline
$\mathsf{H}$ gate & 95.97 & 99.63 & 0.95 \\
$\mathsf{T}$ gate & 99.25 & 99.84 & 0.99 \\
$\mathsf{CZ}$ gate & 98.18 & 99.70 & 0.98 \\
Encoding & 95.78 & 99.74 & 0.95 \\
Decoding & 93.77 & 99.56 & 0.92 \\
Parity (code) & / & 99.86 & 0.98 \\
Parity (error) & / & 99.88 & 0.94 \\
QEC (code) & 92.41 & 99.75 & 0.89 \\
QEC (error) & 96.53 & 99.90 & 0.95 \\
\end{tabular}
\end{ruledtabular}
\label{Table1}
\end{table}

Figure~\ref{fig2} shows examples of gates on bosonic logical qubits. The gate performances are characterized by the Pauli transfer matrix, and Figs.~\ref{fig2}(a) and \ref{fig2}(b) display the deviations of the simulated matrix elements from ideal values~\cite{Gambetta2012PRL_PTM} for the logical $\mathsf{T}$ and Hadamard ($\mathsf{H}$) gates, respectively. The corresponding control pulse duration is set to $2\,\mathrm{\mu s}$, with the optimized waveform for the logical $\mathsf{H}$ gate shown in Fig.~\ref{fig2}(c). The ED universal control is also applicable to multiple logical qubit gates. As an example, we validate an ED logical controlled-phase ($\mathsf{CZ}$) gate between two binomial-encoded logical qubits. Figure~\ref{fig2}(d) shows the differences between the simulated and ideal Pauli transfer matrices for the logical $\mathsf{CZ}$ gate (see Supplementary Materials~\cite{SI} for details). Notably, in contrast to previous ED entangling-gate proposals that require a tunable beam-splitter interaction between two bosonic modes and a dispersively coupled ancilla~\cite{Tsunoda2023PRXQ}, our construction employs only a single dispersively coupled ED ancilla. Comparing the results with and without post-selection (PS), it is confirmed that our ED approach can significantly suppress the imperfections for all logical qubit operations.

Table~\ref{Table1} summarizes the gate performances by systematically comparing ancilla control using $\{\ket{g}, \ket{e}\}$ states, ED control using $\{\ket{g},\ket{f}\}$ states without PS, and ED control with PS. The PS success probability $P_\mathrm{succ}$ represents the average over all linear independent logical states $\{\ket{0_L},\ket{1_L},(\ket{0_L}+\ket{1_L})/\sqrt{2},(\ket{0_L}-i\ket{1_L})/\sqrt{2}\}$. In general, the replacement of the ancilla energy levels by the ED scheme does not substantially degrade the operation fidelity even without PS, indicating that the two-photon drive mechanism provides comparable control quality. More importantly, applying PS dramatically suppresses operation errors to $\epsilon_\mathrm{op}<0.4\%$ for all logical operations, while maintaining high success probabilities around $95\%$. This order-of-magnitude reduction in $\epsilon_\mathrm{op}$ enables the realization of high QEC gains without updating the experimental devices, as indicated by Eq.~(\ref{Eq1}).

\begin{figure}
\centering{}\includegraphics[width=\columnwidth]{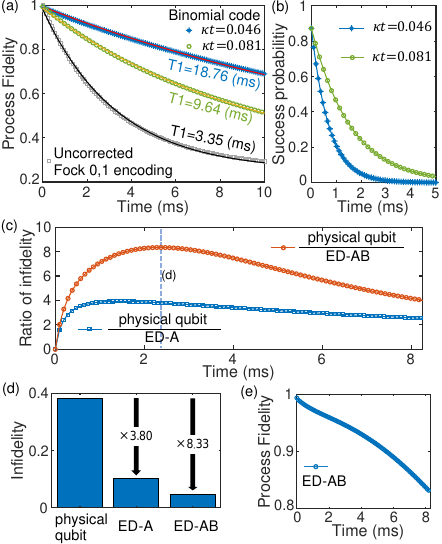}
\caption{Performance of ED QEC processes of a logical qubit. (a) Process fidelity as a function of time for a physical qubit (gray squares) and a binomial code (blue stars and green circles) with repetitive ED-A QEC (error detection on ancilla only). The solid lines represent the fitting curves of the physical qubit (black) and the binomial code (red and green). 
(b) Success probability as a function of time for the binomial code with repetitive ED QEC processes. The blue stars and the green circles represent ED QEC cycle times of $ t=0.046/\kappa$ and $t=0.081/\kappa$, respectively. (c) Process infidelity ratio vs time. The blue curve is the ratio between the physical qubit and an ED-A QEC protected binomial code. The red curve is the ratio between the physical qubit and the binomial code protected by the ED-AB QEC (error detection on both the ancilla and the bosonic mode). (d) Process infidelities corresponding to (c) at 2375~$\mu$s. (e) Process fidelity vs time for an ED-AB QEC protected binomial code.}
\label{fig3}
\end{figure}

\smallskip{}
\noindent\textit{ED QEC of a binomial logical qubit.-} The ED control framework developed above can be directly applied to the QEC cycles to protect a logical qubit. As shown in Table~\ref{Table1}, the ED approach with PS can reduce $\epsilon_{\mathrm{op}}$ by almost one order of magnitude compared to conventional control, implying a significant improvement of $G_{\mathrm{break}}$ to approach $10\times$ beyond break-even.

To validate this prediction, we simulate the evolution of a logical qubit under repetitive QEC cycles using different strategies, as shown in Fig.~\ref{fig3}. In the first strategy, termed ``ED-A" (error detection on ancilla only), we projectively measure the ED ancilla after each operation and discard the trajectories with an outcome of $\ket{e}$. Figure~\ref{fig3}(a) shows the resulting process fidelity as a function of the evolution time. Fitting the decay to $F(t)=Ae^{-t/T_1}+0.25$ yields a logical process fidelity decay time $T_1$ of $18.76\,\mathrm{ms}$ under repeated ED-A QEC cycles. For comparison,  a physical qubit encoded in the Fock basis $\{\ket{0},\ket{1}\}$ without QEC-protection exhibits a coherence time of only $3.35\,\mathrm{ms}$. This corresponds to a gain $G_{\mathrm{break}}=5.60$, representing a $5.60\times$ improvement beyond the break-even point. The heralded nature of ED QEC introduces a trade-off: the success probability decreases with the total number of QEC cycles as failed trajectories accumulate, as shown in Fig.~\ref{fig3}(b). We study two different cycle durations ($t_\mathrm{int}$), showing an {enhanced performance with carefully selected $t_\mathrm{int}$} according to Eq.~(\ref{Eq1}), while a longer $t_\mathrm{int}$ with less frequent PS leads to slower decay of $P_\mathrm{succ}$.

To further improve the QEC gain, we adapt the second strategy, which termed as ``ED-AB" (error detection on both the ancilla and the bosonic mode). As shown in Table~\ref{Table1}, the parity measurement for the error syndrome detection~\cite{SunNature} of the binomially encoded logical qubit has even lower $\epsilon_{\mathrm{op}}$. Therefore, we can reduce the operation error further by post-selecting the error syndrome detection on the logical qubit. Here, to decrease the uncorrectable second-order photon loss errors of the bosonic mode and the first-order errors introduced by the QEC processes, we implement ED QEC only on the code space to correct no-jump evolution errors after three consecutive error syndrome detections and discard trajectories that indicate a single-photon-loss event via the parity measurement.

Figure~\ref{fig3}(c) compares the performances of the logical qubit under the two strategies, where we quantify the improvement using the process infidelity ratio, defined as the physical infidelity divided by the logical infidelity at the same evolution time. For ED-A (blue curve), the logical infidelity is suppressed by a factor of 3.80 relative to the physical baseline at $t=2375\,\mathrm{\mu s}$. In contrast, ED-AB achieves a peak process infidelity ratio of $8.33$ [Fig.~\ref{fig3}(d)].
Figure~\ref{fig3}(e) shows the corresponding process fidelity versus time for ED-AB QEC. The overall evolution deviates slightly from a pure exponential at long times, consistent with the accumulation of a small coherent (systematic) error in the repeated parity measurements (see ~\cite{SI} for further details and discussions).

\begin{figure}
\centering{}\includegraphics[width=\columnwidth]{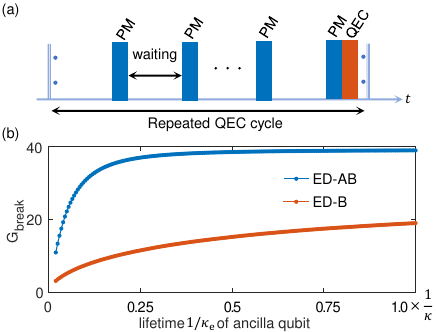}
\caption{Performance of ED QEC processes. (a) Schematic diagram of a single ED QEC cycle. (b) ED QEC gain $G_{\mathrm{break}}$ versus ancilla lifetime $1/\kappa_{\mathrm{e}}$. The blue and red dotted lines represent post-selecting both the ancilla and the bosonic mode (AB) and only post-selecting the bosonic mode (B), respectively. Each point represents the maximum QEC gain $G_{\mathrm{break}}$ for a given lifetime of the ancilla, obtained by optimizing over the interval time $t_{\mathrm{int}}$ and the number of parity measurements $N_{\mathrm{PM}}$.}
\label{fig4}
\end{figure}

\smallskip{}
\noindent\textit{ED QEC gain analysis.-} The simplified model of Eq.~\ref{Eq1} captures how $\epsilon_{\mathrm{op]}}$ and QEC interval $t_\mathrm{int}$ jointly determine the achievable QEC gain. However, when the combined error-detection on both ancilla and bosonic mode are introduced, the analysis of the achievable QEC gain becomes complicated. To understand the fundamental limitations of the ED-AB protocol and identify pathways to further improvement, we develop a compact error budget for a single QEC interval. As illustrated in Fig.~\ref{fig4}(a), each cycle consists of $N_{\mathrm{PM}}$ ED parity measurements (duration $t_{\mathrm{PM}}$ with sub-interval $t_w$), followed by one ED QEC operation (duration
$t_{\mathrm{QEC}}$) that corrects no-jump evolution errors in the code space. Thus, the
total interval of a single QEC cycle is thus $t_{\mathrm{int}} = N_{\mathrm{PM}}(t_w + t_{\mathrm{PM}}) +
t_{\mathrm{QEC}}$. We characterize the QEC gain $G_{\mathrm{break}}$ approximated by the ratio between the physical-qubit infidelity accumulation and the logical infidelity per QEC cycle (see~\cite{SI} for definitions and fitting procedures). In the regime where first-order ancilla damping and single-photon-loss trajectories are detected and discarded, we obtain the approximate scaling. By comparing the infidelity of the physical qubit and infidelity of the ED-AB QEC process in an entire QEC cycle with an interval time $t_{\mathrm{int}}$, we obtain:
\begin{align}
G_{\mathrm{break}}\approx\frac{\alpha\kappa t_{\mathrm{int}}}{N_{\mathrm{PM}}(E_{\mathrm{W}}+E_{\mathrm{PM}})+E_{\mathrm{QEC}}+(\kappa t_{\mathrm{int}})^3},
\label{eq:Gbreak}
\end{align}
where $\alpha=0.6$ represents the ratio of the lifetime of the physical qubit and the break-even point. {The denominator collects the residual errors: $E_{\mathrm{W}}$
from waiting periods, $E_{\mathrm{PM}}$ from parity measurements, $E_{\mathrm{QEC}}$ from the QEC operation, and a higher-order cavity-loss term $(\kappa t_{\mathrm{int}})^3$ representing uncorrectable multi-photon events that accumulate over the full interval.} Substituting the parameters used in Fig.~\ref{fig3}(c) yields $G_{\mathrm{break}}\approx9.33$, in good agreement with the simulated peak gain of $8.33$ shown in Fig~\ref{fig3}(d).

The individual error contributions have apparent physical origins. During waiting, single-photon-loss events are detected by subsequent parity checks and discarded,leaving residual higher-order loss processes and no-jump distortions. This leads to
\begin{align}
&E_{\mathrm{W}}\approx\alpha_{\mathrm{W}}(\kappa t_w)^2,
\label{eq:F_waiting}
\end{align}
with a coefficient $\alpha_{\mathrm{W}}$ determined by the code photon-number distribution (see~\cite{SI}).

For each ED parity measurement, PS removes first-order ancilla and cavity loss trajectories. Therefore, the leading contribution comes form higher-order ancilla damping during the measurement together with residual cavity loss:
\begin{align}
E_{\mathrm{PM}}\approx \alpha_{\mathrm{PMQ}}\kappa_{\mathrm{f}}\kappa_{\mathrm{e}}t_{\mathrm{PM}}^2+\alpha_{\mathrm{PM}}(\kappa t_{\mathrm{PM}})^2,
\label{eq:F_PM}
\end{align}
where the coefficients $\alpha_{\mathrm{PMQ}}$ and $\alpha_{\mathrm{PM}}$ depend on the average energy level populations during the measurement (see~\cite{SI}).

Finally, the ED QEC step contains a higher-order ancilla contribution as well as a residual first-order cavity-loss term that cannot be fully eliminated by PS within the code space, giving
\begin{align}
E_{\mathrm{QEC}}\approx (\alpha_{\mathrm{QEC}} \kappa_{\mathrm{f}}\kappa_{\mathrm{e}}t_{\mathrm{QEC}}^2 +\bar{n} \kappa t_{\mathrm{QEC}})/4.
\label{eq:F_QEC}
\end{align}
Here, the coefficient $\alpha_\mathrm{QEC}$ is related to the mean population of the state $\ket{f}$ during the QEC process and $\bar{n}=2$ for the binomial code. The factor $1/4$ is a rough estimate, with a detailed explanation provided in Supplementary Materials~\cite{SI}.

Equations~(\ref{eq:F_waiting}-\ref{eq:F_QEC}) show that improving the ancilla lifetime rapidly suppresses the dominant ancilla-induced terms in $E_{\mathrm{PM}}$ and $E_{\mathrm{QEC}}$, while the remaining limitations are increasingly set by cavity loss processes during waiting and correction. This competition produces the crossover behavior observed in Fig.~\ref{fig4}(b), where $G_{\mathrm{break}}$ grows quickly with ancilla lifetime up to a critical regime. Beyond the regime, further lifetime improvements yield diminishing returns as remaining cavity loss contributions dominate. For comparison, Fig.~\ref{fig4}(b) also shows an alternative process, ED-B QEC, where the ancilla qubit consists of states $\ket{g}$ and $\ket{e}$ and PS is applied only to the bosonic mode through the parity measurement (See \cite{SI} for more details).

To obtain the critical point, we deduce the relation between the critical lifetime of the ancilla and the bosonic mode lifetime as $1/\kappa_{\mathrm{e}}^{\mathrm{C}}\approx C/\kappa^{\frac{2}{3}}$ with a constant $C=(1\mathrm{\mu s})^{\frac{1}{3}}/1.74$ setting the characteristic time of $1\mathrm{\mu s}$. Here $1/\kappa_{\mathrm{e}}^{\mathrm{C}}$ is the critical lifetime of state $\ket{e}$ of the ancilla. At $\kappa_{\mathrm{e}}=\kappa_{\mathrm{e}}^{\mathrm{C}}$, $G_{\mathrm{break}}$ enters an inflection point. Beyond this point, further increases in the ancilla lifetime do not rapidly increase $G_{\mathrm{break}}$, which instead gradually saturates to $G_{\mathrm{break}}^{\mathrm{sat}}$. The saturation point $G_{\mathrm{break}}^{\mathrm{sat}}\approx 1/[3.8(\tilde{\kappa})+3.8(\tilde{\kappa})^{\frac{2}{3}}]$ is deduced form Eq.~(\ref{eq:Gbreak}) with $\tilde{\kappa}=(1\mathrm{\mu s}) \kappa$ being a dimensionless variable. The analytic expressions for $\kappa_{\mathrm{e}}^{\mathrm{C}}$ and $G_{\mathrm{break}}^{\mathrm{sat}}$ are given in~\cite{SI}. As a concrete example, for $1/\kappa$=2~ms (parameters of the ED-AB QEC in Fig.~\ref{fig4}b) we obtain 1/$\kappa_{\mathrm{e}}^{\mathrm{C}}=91\mathrm{\mu s}$, where the gain is near its optimum $G_{\mathrm{break}}\approx21$, while the predicted saturation is $G_{\mathrm{break}}^{\mathrm{sat}}\approx39$. Increasing the ancilla lifetime from $91\,\mathrm{\mu s}$ to $2\,\mathrm{ms}$ yields only a moderate improvement from 21 to 39, consistent with the numerical trend in Fig.~\ref{fig4}(b).
The saturation originates from residual first-order cavity loss during the correction step [e.g., the $\bar{n} \kappa t_{\mathrm{QEC}}$ term in Eq.~(\ref{eq:F_QEC})].

\smallskip{}
\noindent\textit{Conclusion.-} We have developed an error-detectable (ED) approach to realize high-fidelity universal control on bosonic modes, thereby improving the performance of bosonic logical qubits. With current experimental parameters, we predict a significant improvement in logical operation fidelity. For a binomial code, the process fidelity of a universal logical gate set exceeds $99.6\%$. Our numerical results show a $5.6\times$ extension of the logical qubit coherence beyond the break-even point. Moreover, by applying error detection to the logical qubit instead of error correction operation, we predict a suppression of the logical qubit infidelity by 8.33. To explore the fundamental physical limitations on the QEC, we establish a relationship between the QEC gain and the lifetime of the ancilla, indicating gains beyond 10 times break-even are achievable with current device parameters, e.g., transmon qubit $T_1\sim 100\,\mathrm{\mu s}$.

To further push the QEC gain toward 100 times break-even point~\cite{chen2023QEC}, we propose replacing displacement-based operations with two-photon drives on the bosonic modes. This approach would suppress the remaining first-order errors to higher order. It is anticipated that by eliminating the dominant first-order errors in both the ancilla and the bosonic mode, we will enter an experimental regime where dephasing and other uncorrectable errors become the main limitation. Exploring these effects to develop early fault-tolerant quantum technologies based on ED universal control presents an exciting direction for future work.


\bigskip{}
\noindent \textbf{Acknowledgements}
\begin{acknowledgments}
\noindent This work was funded by the National Natural Science Foundation of China (Grants No. 92265210, 12550006, 92365301, 92165209, 92565301, 12547179 and 12574539) and the Quantum Science and Technology-National Science and Technology Major Project (2021ZD0300200). This work is also supported by the Fundamental Research Funds for the Central Universities, the USTC Research Funds of the Double First-Class Initiative, the supercomputing system in the Supercomputing Center of USTC the USTC Center for Micro and Nanoscale Research and Fabrication.
\end{acknowledgments}


\begin{thebibliography}{39}%
\makeatletter
\providecommand \@ifxundefined [1]{%
 \@ifx{#1\undefined}
}%
\providecommand \@ifnum [1]{%
 \ifnum #1\expandafter \@firstoftwo
 \else \expandafter \@secondoftwo
 \fi
}%
\providecommand \@ifx [1]{%
 \ifx #1\expandafter \@firstoftwo
 \else \expandafter \@secondoftwo
 \fi
}%
\providecommand \natexlab [1]{#1}%
\providecommand \enquote  [1]{``#1''}%
\providecommand \bibnamefont  [1]{#1}%
\providecommand \bibfnamefont [1]{#1}%
\providecommand \citenamefont [1]{#1}%
\providecommand \href@noop [0]{\@secondoftwo}%
\providecommand \href [0]{\begingroup \@sanitize@url \@href}%
\providecommand \@href[1]{\@@startlink{#1}\@@href}%
\providecommand \@@href[1]{\endgroup#1\@@endlink}%
\providecommand \@sanitize@url [0]{\catcode `\\12\catcode `\$12\catcode
  `\&12\catcode `\#12\catcode `\^12\catcode `\_12\catcode `\%12\relax}%
\providecommand \@@startlink[1]{}%
\providecommand \@@endlink[0]{}%
\providecommand \url  [0]{\begingroup\@sanitize@url \@url }%
\providecommand \@url [1]{\endgroup\@href {#1}{\urlprefix }}%
\providecommand \urlprefix  [0]{URL }%
\providecommand \Eprint [0]{\href }%
\providecommand \doibase [0]{http://dx.doi.org/}%
\providecommand \selectlanguage [0]{\@gobble}%
\providecommand \bibinfo  [0]{\@secondoftwo}%
\providecommand \bibfield  [0]{\@secondoftwo}%
\providecommand \translation [1]{[#1]}%
\providecommand \BibitemOpen [0]{}%
\providecommand \bibitemStop [0]{}%
\providecommand \bibitemNoStop [0]{.\EOS\space}%
\providecommand \EOS [0]{\spacefactor3000\relax}%
\providecommand \BibitemShut  [1]{\csname bibitem#1\endcsname}%
\let\auto@bib@innerbib\@empty
\bibitem [{\citenamefont {Nielsen}\ and\ \citenamefont
  {Chuang}(2000)}]{Nielsen}%
  \BibitemOpen
  \bibfield  {author} {\bibinfo {author} {\bibfnamefont {M.~A.}\ \bibnamefont
  {Nielsen}}\ and\ \bibinfo {author} {\bibfnamefont {I.~L.}\ \bibnamefont
  {Chuang}},\ }\href@noop {} {\emph {\bibinfo {title} {Quantum Computation and
  Quantum Information}}}\ (\bibinfo  {publisher} {Cambridge Univ. Press},\
  \bibinfo {year} {2000})\BibitemShut {NoStop}%
\bibitem [{\citenamefont {Shor}(1995)}]{Shor1995}%
  \BibitemOpen
  \bibfield  {author} {\bibinfo {author} {\bibfnamefont {P.~W.}\ \bibnamefont
  {Shor}},\ }\bibfield  {title} {\enquote {\bibinfo {title} {Scheme for
  reducing decoherence in quantum computer memory},}\ }\href {\doibase
  10.1103/PhysRevA.52.R2493} {\bibfield  {journal} {\bibinfo  {journal} {Phys.
  Rev. A}\ }\textbf {\bibinfo {volume} {52}},\ \bibinfo {pages} {R2493}
  (\bibinfo {year} {1995})}\BibitemShut {NoStop}%
\bibitem [{\citenamefont {Shor}(1996)}]{Shor1996FT}%
  \BibitemOpen
  \bibfield  {author} {\bibinfo {author} {\bibfnamefont {P.}~\bibnamefont
  {Shor}},\ }\href {\doibase 10.1109/SFCS.1996.548464} {\emph {\bibinfo {title}
  {Proceedings of 37th Conference on Foundations of Computer Science}}}\
  (\bibinfo {year} {1996})\ pp.\ \bibinfo {pages} {56--65}\BibitemShut
  {NoStop}%
\bibitem [{\citenamefont {DiVincenzo}\ and\ \citenamefont
  {Shor}(1996)}]{DiVincenzoPRL1996}%
  \BibitemOpen
  \bibfield  {author} {\bibinfo {author} {\bibfnamefont {D.~P.}\ \bibnamefont
  {DiVincenzo}}\ and\ \bibinfo {author} {\bibfnamefont {P.~W.}\ \bibnamefont
  {Shor}},\ }\bibfield  {title} {\enquote {\bibinfo {title} {Fault-tolerant
  error correction with efficient quantum codes},}\ }\href {\doibase
  10.1103/PhysRevLett.77.3260} {\bibfield  {journal} {\bibinfo  {journal}
  {Phys. Rev. Lett.}\ }\textbf {\bibinfo {volume} {77}},\ \bibinfo {pages}
  {3260} (\bibinfo {year} {1996})}\BibitemShut {NoStop}%
\bibitem [{\citenamefont {Gottesman}(1998)}]{GottesmanPRA1998}%
  \BibitemOpen
  \bibfield  {author} {\bibinfo {author} {\bibfnamefont {D.}~\bibnamefont
  {Gottesman}},\ }\bibfield  {title} {\enquote {\bibinfo {title} {Theory of
  fault-tolerant quantum computation},}\ }\href {\doibase
  10.1103/PhysRevA.57.127} {\bibfield  {journal} {\bibinfo  {journal} {Phys.
  Rev. A}\ }\textbf {\bibinfo {volume} {57}},\ \bibinfo {pages} {127} (\bibinfo
  {year} {1998})}\BibitemShut {NoStop}%
\bibitem [{\citenamefont {PRESKILL}(1998)}]{preskill1998}%
  \BibitemOpen
  \bibfield  {author} {\bibinfo {author} {\bibfnamefont {J.}~\bibnamefont
  {PRESKILL}},\ }\href {\doibase 10.1142/9789812385253_0008} {\emph {\bibinfo
  {title} {Introduction to Quantum Computation and Information}}}\ (\bibinfo
  {year} {1998})\ pp.\ \bibinfo {pages} {213--269}\BibitemShut {NoStop}%
\bibitem [{\citenamefont {Kitaev}(2003)}]{KITAEV20032}%
  \BibitemOpen
  \bibfield  {author} {\bibinfo {author} {\bibfnamefont {A.}~\bibnamefont
  {Kitaev}},\ }\bibfield  {title} {\enquote {\bibinfo {title} {Fault-tolerant
  quantum computation by anyons},}\ }\href {\doibase
  https://doi.org/10.1016/S0003-4916(02)00018-0} {\bibfield  {journal}
  {\bibinfo  {journal} {Annals of Physics}\ }\textbf {\bibinfo {volume}
  {303}},\ \bibinfo {pages} {2} (\bibinfo {year} {2003})}\BibitemShut {NoStop}%
\bibitem [{\citenamefont {Gidney}\ and\ \citenamefont
  {Eker{\aa{}}}(2021)}]{Gidney2021howtofactorbit}%
  \BibitemOpen
  \bibfield  {author} {\bibinfo {author} {\bibfnamefont {C.}~\bibnamefont
  {Gidney}}\ and\ \bibinfo {author} {\bibfnamefont {M.}~\bibnamefont
  {Eker{\aa{}}}},\ }\bibfield  {title} {\enquote {\bibinfo {title} {How to
  factor 2048 bit {RSA} integers in 8 hours using 20 million noisy qubits},}\
  }\href {\doibase 10.22331/q-2021-04-15-433} {\bibfield  {journal} {\bibinfo
  {journal} {{Quantum}}\ }\textbf {\bibinfo {volume} {5}},\ \bibinfo {pages}
  {433} (\bibinfo {year} {2021})}\BibitemShut {NoStop}%
\bibitem [{\citenamefont {Gidney}(2025)}]{gidney2025factor2048bitrsa}%
  \BibitemOpen
  \bibfield  {author} {\bibinfo {author} {\bibfnamefont {C.}~\bibnamefont
  {Gidney}},\ }\bibfield  {title} {\enquote {\bibinfo {title} {How to factor
  2048 bit rsa integers with less than a million noisy qubits},}\ }\href
  {https://arxiv.org/abs/2505.15917} {\bibfield  {journal} {\bibinfo  {journal}
  {arXiv: 2505.15917}\ } (\bibinfo {year} {2025})}\BibitemShut {NoStop}%
\bibitem [{\citenamefont {Ofek}\ \emph {et~al.}(2016)\citenamefont {Ofek},
  \citenamefont {Petrenko}, \citenamefont {Heeres}, \citenamefont {Reinhold},
  \citenamefont {Leghtas}, \citenamefont {Vlastakis}, \citenamefont {Liu},
  \citenamefont {Frunzio}, \citenamefont {Girvin}, \citenamefont {Jiang},
  \citenamefont {Mirrahimi}, \citenamefont {Devoret},\ and\ \citenamefont
  {Schoelkopf}}]{Ofek2016}%
  \BibitemOpen
  \bibfield  {author} {\bibinfo {author} {\bibfnamefont {N.}~\bibnamefont
  {Ofek}}, \bibinfo {author} {\bibfnamefont {A.}~\bibnamefont {Petrenko}},
  \bibinfo {author} {\bibfnamefont {R.}~\bibnamefont {Heeres}}, \bibinfo
  {author} {\bibfnamefont {P.}~\bibnamefont {Reinhold}}, \bibinfo {author}
  {\bibfnamefont {Z.}~\bibnamefont {Leghtas}}, \bibinfo {author} {\bibfnamefont
  {B.}~\bibnamefont {Vlastakis}}, \bibinfo {author} {\bibfnamefont
  {Y.}~\bibnamefont {Liu}}, \bibinfo {author} {\bibfnamefont {L.}~\bibnamefont
  {Frunzio}}, \bibinfo {author} {\bibfnamefont {S.~M.}\ \bibnamefont {Girvin}},
  \bibinfo {author} {\bibfnamefont {L.}~\bibnamefont {Jiang}}, \bibinfo
  {author} {\bibfnamefont {M.}~\bibnamefont {Mirrahimi}}, \bibinfo {author}
  {\bibfnamefont {M.~H.}\ \bibnamefont {Devoret}}, \ and\ \bibinfo {author}
  {\bibfnamefont {R.~J.}\ \bibnamefont {Schoelkopf}},\ }\bibfield  {title}
  {\enquote {\bibinfo {title} {Extending the lifetime of a quantum bit with
  error correction in superconducting circuits},}\ }\href {\doibase
  10.1038/nature18949} {\bibfield  {journal} {\bibinfo  {journal} {Nature}\
  }\textbf {\bibinfo {volume} {536}},\ \bibinfo {pages} {441} (\bibinfo {year}
  {2016})}\BibitemShut {NoStop}%
\bibitem [{\citenamefont {Ni}\ \emph {et~al.}(2023)\citenamefont {Ni},
  \citenamefont {Li}, \citenamefont {Deng}, \citenamefont {Cai}, \citenamefont
  {Zhang}, \citenamefont {Wang}, \citenamefont {Yang}, \citenamefont {Yu},
  \citenamefont {Yan}, \citenamefont {Liu}, \citenamefont {Zou}, \citenamefont
  {Sun}, \citenamefont {Zheng}, \citenamefont {Xu},\ and\ \citenamefont
  {Yu}}]{Ni2023Nature}%
  \BibitemOpen
  \bibfield  {author} {\bibinfo {author} {\bibfnamefont {Z.}~\bibnamefont
  {Ni}}, \bibinfo {author} {\bibfnamefont {S.}~\bibnamefont {Li}}, \bibinfo
  {author} {\bibfnamefont {X.}~\bibnamefont {Deng}}, \bibinfo {author}
  {\bibfnamefont {Y.}~\bibnamefont {Cai}}, \bibinfo {author} {\bibfnamefont
  {L.}~\bibnamefont {Zhang}}, \bibinfo {author} {\bibfnamefont
  {W.}~\bibnamefont {Wang}}, \bibinfo {author} {\bibfnamefont {Z.-B.}\
  \bibnamefont {Yang}}, \bibinfo {author} {\bibfnamefont {H.}~\bibnamefont
  {Yu}}, \bibinfo {author} {\bibfnamefont {F.}~\bibnamefont {Yan}}, \bibinfo
  {author} {\bibfnamefont {S.}~\bibnamefont {Liu}}, \bibinfo {author}
  {\bibfnamefont {C.-L.}\ \bibnamefont {Zou}}, \bibinfo {author} {\bibfnamefont
  {L.}~\bibnamefont {Sun}}, \bibinfo {author} {\bibfnamefont {S.-B.}\
  \bibnamefont {Zheng}}, \bibinfo {author} {\bibfnamefont {Y.}~\bibnamefont
  {Xu}}, \ and\ \bibinfo {author} {\bibfnamefont {D.}~\bibnamefont {Yu}},\
  }\bibfield  {title} {\enquote {\bibinfo {title} {Beating the break-even point
  with a discrete-variable-encoded logical qubit},}\ }\href {\doibase
  10.1038/s41586-023-05784-4} {\bibfield  {journal} {\bibinfo  {journal}
  {Nature}\ }\textbf {\bibinfo {volume} {616}},\ \bibinfo {pages} {56}
  (\bibinfo {year} {2023})}\BibitemShut {NoStop}%
\bibitem [{\citenamefont {Sivak}\ \emph {et~al.}(2023)\citenamefont {Sivak},
  \citenamefont {Eickbusch}, \citenamefont {Royer}, \citenamefont {Singh},
  \citenamefont {Tsioutsios}, \citenamefont {Ganjam}, \citenamefont {Miano},
  \citenamefont {Brock}, \citenamefont {Ding}, \citenamefont {Frunzio},
  \citenamefont {Girvin}, \citenamefont {Schoelkopf},\ and\ \citenamefont
  {Devoret}}]{Sivak2023}%
  \BibitemOpen
  \bibfield  {author} {\bibinfo {author} {\bibfnamefont {V.~V.}\ \bibnamefont
  {Sivak}}, \bibinfo {author} {\bibfnamefont {A.}~\bibnamefont {Eickbusch}},
  \bibinfo {author} {\bibfnamefont {B.}~\bibnamefont {Royer}}, \bibinfo
  {author} {\bibfnamefont {S.}~\bibnamefont {Singh}}, \bibinfo {author}
  {\bibfnamefont {I.}~\bibnamefont {Tsioutsios}}, \bibinfo {author}
  {\bibfnamefont {S.}~\bibnamefont {Ganjam}}, \bibinfo {author} {\bibfnamefont
  {A.}~\bibnamefont {Miano}}, \bibinfo {author} {\bibfnamefont {B.~L.}\
  \bibnamefont {Brock}}, \bibinfo {author} {\bibfnamefont {A.~Z.}\ \bibnamefont
  {Ding}}, \bibinfo {author} {\bibfnamefont {L.}~\bibnamefont {Frunzio}},
  \bibinfo {author} {\bibfnamefont {S.~M.}\ \bibnamefont {Girvin}}, \bibinfo
  {author} {\bibfnamefont {R.~J.}\ \bibnamefont {Schoelkopf}}, \ and\ \bibinfo
  {author} {\bibfnamefont {M.~H.}\ \bibnamefont {Devoret}},\ }\bibfield
  {title} {\enquote {\bibinfo {title} {Real-time quantum error correction
  beyond break-even},}\ }\href {\doibase 10.1038/s41586-023-05782-6} {\bibfield
   {journal} {\bibinfo  {journal} {Nature}\ }\textbf {\bibinfo {volume}
  {616}},\ \bibinfo {pages} {50} (\bibinfo {year} {2023})}\BibitemShut
  {NoStop}%
\bibitem [{\citenamefont {Acharya}\ and\ \citenamefont
  {et~al.}(2025)}]{Google2024}%
  \BibitemOpen
  \bibfield  {author} {\bibinfo {author} {\bibfnamefont {R.}~\bibnamefont
  {Acharya}}\ and\ \bibinfo {author} {\bibnamefont {et~al.}},\ }\bibfield
  {title} {\enquote {\bibinfo {title} {Quantum error correction below the
  surface code threshold},}\ }\href {\doibase 10.1038/s41586-024-08449-y}
  {\bibfield  {journal} {\bibinfo  {journal} {Nature}\ }\textbf {\bibinfo
  {volume} {638}},\ \bibinfo {pages} {920} (\bibinfo {year}
  {2025})}\BibitemShut {NoStop}%
\bibitem [{\citenamefont {Chen}\ \emph {et~al.}(2023)\citenamefont {Chen},
  \citenamefont {Sun},\ and\ \citenamefont {Zou}}]{chen2023QEC}%
  \BibitemOpen
  \bibfield  {author} {\bibinfo {author} {\bibfnamefont {Z.}~\bibnamefont
  {Chen}}, \bibinfo {author} {\bibfnamefont {L.}~\bibnamefont {Sun}}, \ and\
  \bibinfo {author} {\bibfnamefont {C.-L.}\ \bibnamefont {Zou}},\ }\bibfield
  {title} {\enquote {\bibinfo {title} {Entering the error-corrected quantum
  era},}\ }\href {\doibase https://doi.org/10.1016/j.scib.2023.04.039}
  {\bibfield  {journal} {\bibinfo  {journal} {Sci. Bull.}\ }\textbf {\bibinfo
  {volume} {68}},\ \bibinfo {pages} {961} (\bibinfo {year} {2023})}\BibitemShut
  {NoStop}%
\bibitem [{\citenamefont {Chuang}\ \emph {et~al.}(1997)\citenamefont {Chuang},
  \citenamefont {Leung},\ and\ \citenamefont {Yamamoto}}]{ChuangPRA1997}%
  \BibitemOpen
  \bibfield  {author} {\bibinfo {author} {\bibfnamefont {I.~L.}\ \bibnamefont
  {Chuang}}, \bibinfo {author} {\bibfnamefont {D.~W.}\ \bibnamefont {Leung}}, \
  and\ \bibinfo {author} {\bibfnamefont {Y.}~\bibnamefont {Yamamoto}},\
  }\bibfield  {title} {\enquote {\bibinfo {title} {Bosonic quantum codes for
  amplitude damping},}\ }\href {\doibase 10.1103/PhysRevA.56.1114} {\bibfield
  {journal} {\bibinfo  {journal} {Phys. Rev. A}\ }\textbf {\bibinfo {volume}
  {56}},\ \bibinfo {pages} {1114} (\bibinfo {year} {1997})}\BibitemShut
  {NoStop}%
\bibitem [{\citenamefont {Braunstein}(1998)}]{BraunsteinPRL1998}%
  \BibitemOpen
  \bibfield  {author} {\bibinfo {author} {\bibfnamefont {S.~L.}\ \bibnamefont
  {Braunstein}},\ }\bibfield  {title} {\enquote {\bibinfo {title} {Error
  correction for continuous quantum variables},}\ }\href {\doibase
  10.1103/PhysRevLett.80.4084} {\bibfield  {journal} {\bibinfo  {journal}
  {Phys. Rev. Lett.}\ }\textbf {\bibinfo {volume} {80}},\ \bibinfo {pages}
  {4084} (\bibinfo {year} {1998})}\BibitemShut {NoStop}%
\bibitem [{\citenamefont {Gottesman}\ \emph {et~al.}(2001)\citenamefont
  {Gottesman}, \citenamefont {Kitaev},\ and\ \citenamefont
  {Preskill}}]{Gottesman2001PRA}%
  \BibitemOpen
  \bibfield  {author} {\bibinfo {author} {\bibfnamefont {D.}~\bibnamefont
  {Gottesman}}, \bibinfo {author} {\bibfnamefont {A.}~\bibnamefont {Kitaev}}, \
  and\ \bibinfo {author} {\bibfnamefont {J.}~\bibnamefont {Preskill}},\
  }\bibfield  {title} {\enquote {\bibinfo {title} {Encoding a qubit in an
  oscillator},}\ }\href {\doibase 10.1103/PhysRevA.64.012310} {\bibfield
  {journal} {\bibinfo  {journal} {Phys. Rev. A}\ }\textbf {\bibinfo {volume}
  {64}},\ \bibinfo {pages} {012310} (\bibinfo {year} {2001})}\BibitemShut
  {NoStop}%
\bibitem [{\citenamefont {Cochrane}\ \emph {et~al.}(1999)\citenamefont
  {Cochrane}, \citenamefont {Milburn},\ and\ \citenamefont
  {Munro}}]{CochranePRA1999}%
  \BibitemOpen
  \bibfield  {author} {\bibinfo {author} {\bibfnamefont {P.~T.}\ \bibnamefont
  {Cochrane}}, \bibinfo {author} {\bibfnamefont {G.~J.}\ \bibnamefont
  {Milburn}}, \ and\ \bibinfo {author} {\bibfnamefont {W.~J.}\ \bibnamefont
  {Munro}},\ }\bibfield  {title} {\enquote {\bibinfo {title} {Macroscopically
  distinct quantum-superposition states as a bosonic code for amplitude
  damping},}\ }\href {\doibase 10.1103/PhysRevA.59.2631} {\bibfield  {journal}
  {\bibinfo  {journal} {Phys. Rev. A}\ }\textbf {\bibinfo {volume} {59}},\
  \bibinfo {pages} {2631} (\bibinfo {year} {1999})}\BibitemShut {NoStop}%
\bibitem [{\citenamefont {Michael}\ \emph {et~al.}(2016)\citenamefont
  {Michael}, \citenamefont {Silveri}, \citenamefont {Brierley}, \citenamefont
  {Albert}, \citenamefont {Salmilehto}, \citenamefont {Jiang},\ and\
  \citenamefont {Girvin}}]{Michael2016PRX}%
  \BibitemOpen
  \bibfield  {author} {\bibinfo {author} {\bibfnamefont {M.~H.}\ \bibnamefont
  {Michael}}, \bibinfo {author} {\bibfnamefont {M.}~\bibnamefont {Silveri}},
  \bibinfo {author} {\bibfnamefont {R.~T.}\ \bibnamefont {Brierley}}, \bibinfo
  {author} {\bibfnamefont {V.~V.}\ \bibnamefont {Albert}}, \bibinfo {author}
  {\bibfnamefont {J.}~\bibnamefont {Salmilehto}}, \bibinfo {author}
  {\bibfnamefont {L.}~\bibnamefont {Jiang}}, \ and\ \bibinfo {author}
  {\bibfnamefont {S.~M.}\ \bibnamefont {Girvin}},\ }\bibfield  {title}
  {\enquote {\bibinfo {title} {New class of quantum error-correcting codes for
  a bosonic mode},}\ }\href {\doibase 10.1103/PhysRevX.6.031006} {\bibfield
  {journal} {\bibinfo  {journal} {Phys. Rev. X}\ }\textbf {\bibinfo {volume}
  {6}},\ \bibinfo {pages} {031006} (\bibinfo {year} {2016})}\BibitemShut
  {NoStop}%
\bibitem [{\citenamefont {Albert}\ \emph {et~al.}(2018)\citenamefont {Albert},
  \citenamefont {Noh}, \citenamefont {Duivenvoorden}, \citenamefont {Young},
  \citenamefont {Brierley}, \citenamefont {Reinhold}, \citenamefont {Vuillot},
  \citenamefont {Li}, \citenamefont {Shen}, \citenamefont {Girvin},
  \citenamefont {Terhal},\ and\ \citenamefont {Jiang}}]{AlbertPRA2018}%
  \BibitemOpen
  \bibfield  {author} {\bibinfo {author} {\bibfnamefont {V.~V.}\ \bibnamefont
  {Albert}}, \bibinfo {author} {\bibfnamefont {K.}~\bibnamefont {Noh}},
  \bibinfo {author} {\bibfnamefont {K.}~\bibnamefont {Duivenvoorden}}, \bibinfo
  {author} {\bibfnamefont {D.~J.}\ \bibnamefont {Young}}, \bibinfo {author}
  {\bibfnamefont {R.~T.}\ \bibnamefont {Brierley}}, \bibinfo {author}
  {\bibfnamefont {P.}~\bibnamefont {Reinhold}}, \bibinfo {author}
  {\bibfnamefont {C.}~\bibnamefont {Vuillot}}, \bibinfo {author} {\bibfnamefont
  {L.}~\bibnamefont {Li}}, \bibinfo {author} {\bibfnamefont {C.}~\bibnamefont
  {Shen}}, \bibinfo {author} {\bibfnamefont {S.~M.}\ \bibnamefont {Girvin}},
  \bibinfo {author} {\bibfnamefont {B.~M.}\ \bibnamefont {Terhal}}, \ and\
  \bibinfo {author} {\bibfnamefont {L.}~\bibnamefont {Jiang}},\ }\bibfield
  {title} {\enquote {\bibinfo {title} {Performance and structure of single-mode
  bosonic codes},}\ }\href {\doibase 10.1103/PhysRevA.97.032346} {\bibfield
  {journal} {\bibinfo  {journal} {Phys. Rev. A}\ }\textbf {\bibinfo {volume}
  {97}},\ \bibinfo {pages} {032346} (\bibinfo {year} {2018})}\BibitemShut
  {NoStop}%
\bibitem [{\citenamefont {Cai}\ \emph {et~al.}(2021)\citenamefont {Cai},
  \citenamefont {Ma}, \citenamefont {Wang}, \citenamefont {Zou},\ and\
  \citenamefont {Sun}}]{cai2020bosonic}%
  \BibitemOpen
  \bibfield  {author} {\bibinfo {author} {\bibfnamefont {W.}~\bibnamefont
  {Cai}}, \bibinfo {author} {\bibfnamefont {Y.}~\bibnamefont {Ma}}, \bibinfo
  {author} {\bibfnamefont {W.}~\bibnamefont {Wang}}, \bibinfo {author}
  {\bibfnamefont {C.-L.}\ \bibnamefont {Zou}}, \ and\ \bibinfo {author}
  {\bibfnamefont {L.}~\bibnamefont {Sun}},\ }\bibfield  {title} {\enquote
  {\bibinfo {title} {Bosonic quantum error correction codes in superconducting
  quantum circuits},}\ }\href {\doibase
  https://doi.org/10.1016/j.fmre.2020.12.006} {\bibfield  {journal} {\bibinfo
  {journal} {Fundamental Research}\ }\textbf {\bibinfo {volume} {1}},\ \bibinfo
  {pages} {50} (\bibinfo {year} {2021})}\BibitemShut {NoStop}%
\bibitem [{\citenamefont {Ma}\ \emph {et~al.}(2021)\citenamefont {Ma},
  \citenamefont {Puri}, \citenamefont {Schoelkopf}, \citenamefont {Devoret},
  \citenamefont {Girvin},\ and\ \citenamefont {Jiang}}]{MA2021Bosonic}%
  \BibitemOpen
  \bibfield  {author} {\bibinfo {author} {\bibfnamefont {W.-L.}\ \bibnamefont
  {Ma}}, \bibinfo {author} {\bibfnamefont {S.}~\bibnamefont {Puri}}, \bibinfo
  {author} {\bibfnamefont {R.~J.}\ \bibnamefont {Schoelkopf}}, \bibinfo
  {author} {\bibfnamefont {M.~H.}\ \bibnamefont {Devoret}}, \bibinfo {author}
  {\bibfnamefont {S.}~\bibnamefont {Girvin}}, \ and\ \bibinfo {author}
  {\bibfnamefont {L.}~\bibnamefont {Jiang}},\ }\bibfield  {title} {\enquote
  {\bibinfo {title} {Quantum control of bosonic modes with superconducting
  circuits},}\ }\href {\doibase https://doi.org/10.1016/j.scib.2021.05.024}
  {\bibfield  {journal} {\bibinfo  {journal} {Sci. Bull.}\ }\textbf {\bibinfo
  {volume} {66}},\ \bibinfo {pages} {1789} (\bibinfo {year}
  {2021})}\BibitemShut {NoStop}%
\bibitem [{\citenamefont {Joshi}\ \emph {et~al.}(2021)\citenamefont {Joshi},
  \citenamefont {Noh},\ and\ \citenamefont {Gao}}]{Joshi2021Bosonic}%
  \BibitemOpen
  \bibfield  {author} {\bibinfo {author} {\bibfnamefont {A.}~\bibnamefont
  {Joshi}}, \bibinfo {author} {\bibfnamefont {K.}~\bibnamefont {Noh}}, \ and\
  \bibinfo {author} {\bibfnamefont {Y.~Y.}\ \bibnamefont {Gao}},\ }\bibfield
  {title} {\enquote {\bibinfo {title} {Quantum information processing with
  bosonic qubits in circuit qed},}\ }\href {\doibase 10.1088/2058-9565/abe989}
  {\bibfield  {journal} {\bibinfo  {journal} {Quantum Science and Technology}\
  }\textbf {\bibinfo {volume} {6}},\ \bibinfo {pages} {033001} (\bibinfo {year}
  {2021})}\BibitemShut {NoStop}%
\bibitem [{\citenamefont {Krastanov}\ \emph {et~al.}(2015)\citenamefont
  {Krastanov}, \citenamefont {Albert}, \citenamefont {Shen}, \citenamefont
  {Zou}, \citenamefont {Heeres}, \citenamefont {Vlastakis}, \citenamefont
  {Schoelkopf},\ and\ \citenamefont {Jiang}}]{Krastanov2015}%
  \BibitemOpen
  \bibfield  {author} {\bibinfo {author} {\bibfnamefont {S.}~\bibnamefont
  {Krastanov}}, \bibinfo {author} {\bibfnamefont {V.~V.}\ \bibnamefont
  {Albert}}, \bibinfo {author} {\bibfnamefont {C.}~\bibnamefont {Shen}},
  \bibinfo {author} {\bibfnamefont {C.-L.}\ \bibnamefont {Zou}}, \bibinfo
  {author} {\bibfnamefont {R.~W.}\ \bibnamefont {Heeres}}, \bibinfo {author}
  {\bibfnamefont {B.}~\bibnamefont {Vlastakis}}, \bibinfo {author}
  {\bibfnamefont {R.~J.}\ \bibnamefont {Schoelkopf}}, \ and\ \bibinfo {author}
  {\bibfnamefont {L.}~\bibnamefont {Jiang}},\ }\bibfield  {title} {\enquote
  {\bibinfo {title} {Universal control of an oscillator with dispersive
  coupling to a qubit},}\ }\href {\doibase 10.1103/PhysRevA.92.040303}
  {\bibfield  {journal} {\bibinfo  {journal} {Phys. Rev. A}\ }\textbf {\bibinfo
  {volume} {92}},\ \bibinfo {pages} {040303} (\bibinfo {year}
  {2015})}\BibitemShut {NoStop}%
\bibitem [{\citenamefont {Heeres}\ \emph {et~al.}(2015)\citenamefont {Heeres},
  \citenamefont {Vlastakis}, \citenamefont {Holland}, \citenamefont
  {Krastanov}, \citenamefont {Albert}, \citenamefont {Frunzio}, \citenamefont
  {Jiang},\ and\ \citenamefont {Schoelkopf}}]{HeeresPRL2015}%
  \BibitemOpen
  \bibfield  {author} {\bibinfo {author} {\bibfnamefont {R.~W.}\ \bibnamefont
  {Heeres}}, \bibinfo {author} {\bibfnamefont {B.}~\bibnamefont {Vlastakis}},
  \bibinfo {author} {\bibfnamefont {E.}~\bibnamefont {Holland}}, \bibinfo
  {author} {\bibfnamefont {S.}~\bibnamefont {Krastanov}}, \bibinfo {author}
  {\bibfnamefont {V.~V.}\ \bibnamefont {Albert}}, \bibinfo {author}
  {\bibfnamefont {L.}~\bibnamefont {Frunzio}}, \bibinfo {author} {\bibfnamefont
  {L.}~\bibnamefont {Jiang}}, \ and\ \bibinfo {author} {\bibfnamefont {R.~J.}\
  \bibnamefont {Schoelkopf}},\ }\bibfield  {title} {\enquote {\bibinfo {title}
  {Cavity state manipulation using photon-number selective phase gates},}\
  }\href {\doibase 10.1103/PhysRevLett.115.137002} {\bibfield  {journal}
  {\bibinfo  {journal} {Phys. Rev. Lett.}\ }\textbf {\bibinfo {volume} {115}},\
  \bibinfo {pages} {137002} (\bibinfo {year} {2015})}\BibitemShut {NoStop}%
\bibitem [{\citenamefont {Heeres}\ \emph {et~al.}(2017)\citenamefont {Heeres},
  \citenamefont {Reinhold}, \citenamefont {Ofek}, \citenamefont {Frunzio},
  \citenamefont {Jiang}, \citenamefont {Devoret},\ and\ \citenamefont
  {Schoelkopf}}]{Heeres2017}%
  \BibitemOpen
  \bibfield  {author} {\bibinfo {author} {\bibfnamefont {R.~W.}\ \bibnamefont
  {Heeres}}, \bibinfo {author} {\bibfnamefont {P.}~\bibnamefont {Reinhold}},
  \bibinfo {author} {\bibfnamefont {N.}~\bibnamefont {Ofek}}, \bibinfo {author}
  {\bibfnamefont {L.}~\bibnamefont {Frunzio}}, \bibinfo {author} {\bibfnamefont
  {L.}~\bibnamefont {Jiang}}, \bibinfo {author} {\bibfnamefont {M.~H.}\
  \bibnamefont {Devoret}}, \ and\ \bibinfo {author} {\bibfnamefont {R.~J.}\
  \bibnamefont {Schoelkopf}},\ }\bibfield  {title} {\enquote {\bibinfo {title}
  {Implementing a universal gate set on a logical qubit encoded in an
  oscillator},}\ }\href {\doibase 10.1038/s41467-017-00045-1} {\bibfield
  {journal} {\bibinfo  {journal} {Nat. Commun.}\ }\textbf {\bibinfo {volume}
  {8}},\ \bibinfo {pages} {94} (\bibinfo {year} {2017})}\BibitemShut {NoStop}%
\bibitem [{\citenamefont {Hu}\ \emph {et~al.}(2019)\citenamefont {Hu},
  \citenamefont {Ma}, \citenamefont {Cai}, \citenamefont {Mu}, \citenamefont
  {Xu}, \citenamefont {Wang}, \citenamefont {Wu}, \citenamefont {Wang},
  \citenamefont {Song}, \citenamefont {Zou}, \citenamefont {Girvin},
  \citenamefont {Duan},\ and\ \citenamefont {Sun}}]{Hu2019}%
  \BibitemOpen
  \bibfield  {author} {\bibinfo {author} {\bibfnamefont {L.}~\bibnamefont
  {Hu}}, \bibinfo {author} {\bibfnamefont {Y.}~\bibnamefont {Ma}}, \bibinfo
  {author} {\bibfnamefont {W.}~\bibnamefont {Cai}}, \bibinfo {author}
  {\bibfnamefont {X.}~\bibnamefont {Mu}}, \bibinfo {author} {\bibfnamefont
  {Y.}~\bibnamefont {Xu}}, \bibinfo {author} {\bibfnamefont {W.}~\bibnamefont
  {Wang}}, \bibinfo {author} {\bibfnamefont {Y.}~\bibnamefont {Wu}}, \bibinfo
  {author} {\bibfnamefont {H.}~\bibnamefont {Wang}}, \bibinfo {author}
  {\bibfnamefont {Y.~P.}\ \bibnamefont {Song}}, \bibinfo {author}
  {\bibfnamefont {C.-L.}\ \bibnamefont {Zou}}, \bibinfo {author} {\bibfnamefont
  {S.~M.}\ \bibnamefont {Girvin}}, \bibinfo {author} {\bibfnamefont {L.-M.}\
  \bibnamefont {Duan}}, \ and\ \bibinfo {author} {\bibfnamefont
  {L.}~\bibnamefont {Sun}},\ }\bibfield  {title} {\enquote {\bibinfo {title}
  {Quantum error correction and universal gate set operation on a binomial
  bosonic logical qubit},}\ }\href {\doibase 10.1038/s41567-018-0414-3}
  {\bibfield  {journal} {\bibinfo  {journal} {Nat. Phys.}\ }\textbf {\bibinfo
  {volume} {15}},\ \bibinfo {pages} {503} (\bibinfo {year} {2019})}\BibitemShut
  {NoStop}%
\bibitem [{\citenamefont {Reagor}\ \emph {et~al.}(2016)\citenamefont {Reagor},
  \citenamefont {Pfaff}, \citenamefont {Axline}, \citenamefont {Heeres},
  \citenamefont {Ofek}, \citenamefont {Sliwa}, \citenamefont {Holland},
  \citenamefont {Wang}, \citenamefont {Blumoff}, \citenamefont {Chou},
  \citenamefont {Hatridge}, \citenamefont {Frunzio}, \citenamefont {Devoret},
  \citenamefont {Jiang},\ and\ \citenamefont {Schoelkopf}}]{ReagorPRB2016}%
  \BibitemOpen
  \bibfield  {author} {\bibinfo {author} {\bibfnamefont {M.}~\bibnamefont
  {Reagor}}, \bibinfo {author} {\bibfnamefont {W.}~\bibnamefont {Pfaff}},
  \bibinfo {author} {\bibfnamefont {C.}~\bibnamefont {Axline}}, \bibinfo
  {author} {\bibfnamefont {R.~W.}\ \bibnamefont {Heeres}}, \bibinfo {author}
  {\bibfnamefont {N.}~\bibnamefont {Ofek}}, \bibinfo {author} {\bibfnamefont
  {K.}~\bibnamefont {Sliwa}}, \bibinfo {author} {\bibfnamefont
  {E.}~\bibnamefont {Holland}}, \bibinfo {author} {\bibfnamefont
  {C.}~\bibnamefont {Wang}}, \bibinfo {author} {\bibfnamefont {J.}~\bibnamefont
  {Blumoff}}, \bibinfo {author} {\bibfnamefont {K.}~\bibnamefont {Chou}},
  \bibinfo {author} {\bibfnamefont {M.~J.}\ \bibnamefont {Hatridge}}, \bibinfo
  {author} {\bibfnamefont {L.}~\bibnamefont {Frunzio}}, \bibinfo {author}
  {\bibfnamefont {M.~H.}\ \bibnamefont {Devoret}}, \bibinfo {author}
  {\bibfnamefont {L.}~\bibnamefont {Jiang}}, \ and\ \bibinfo {author}
  {\bibfnamefont {R.~J.}\ \bibnamefont {Schoelkopf}},\ }\bibfield  {title}
  {\enquote {\bibinfo {title} {Quantum memory with millisecond coherence in
  circuit qed},}\ }\href {\doibase 10.1103/PhysRevB.94.014506} {\bibfield
  {journal} {\bibinfo  {journal} {Phys. Rev. B}\ }\textbf {\bibinfo {volume}
  {94}},\ \bibinfo {pages} {014506} (\bibinfo {year} {2016})}\BibitemShut
  {NoStop}%
\bibitem [{\citenamefont {Milul}\ \emph {et~al.}(2023)\citenamefont {Milul},
  \citenamefont {Guttel}, \citenamefont {Goldblatt}, \citenamefont {Hazanov},
  \citenamefont {Joshi}, \citenamefont {Chausovsky}, \citenamefont {Kahn},
  \citenamefont {\ifmmode~\mbox{\c{C}}\else \c{C}\fi{}ifty\"urek},
  \citenamefont {Lafont},\ and\ \citenamefont {Rosenblum}}]{Milul2023PRXQ}%
  \BibitemOpen
  \bibfield  {author} {\bibinfo {author} {\bibfnamefont {O.}~\bibnamefont
  {Milul}}, \bibinfo {author} {\bibfnamefont {B.}~\bibnamefont {Guttel}},
  \bibinfo {author} {\bibfnamefont {U.}~\bibnamefont {Goldblatt}}, \bibinfo
  {author} {\bibfnamefont {S.}~\bibnamefont {Hazanov}}, \bibinfo {author}
  {\bibfnamefont {L.~M.}\ \bibnamefont {Joshi}}, \bibinfo {author}
  {\bibfnamefont {D.}~\bibnamefont {Chausovsky}}, \bibinfo {author}
  {\bibfnamefont {N.}~\bibnamefont {Kahn}}, \bibinfo {author} {\bibfnamefont
  {E.}~\bibnamefont {\ifmmode~\mbox{\c{C}}\else \c{C}\fi{}ifty\"urek}},
  \bibinfo {author} {\bibfnamefont {F.}~\bibnamefont {Lafont}}, \ and\ \bibinfo
  {author} {\bibfnamefont {S.}~\bibnamefont {Rosenblum}},\ }\bibfield  {title}
  {\enquote {\bibinfo {title} {Superconducting cavity qubit with tens of
  milliseconds single-photon coherence time},}\ }\href {\doibase
  10.1103/PRXQuantum.4.030336} {\bibfield  {journal} {\bibinfo  {journal} {PRX
  Quantum}\ }\textbf {\bibinfo {volume} {4}},\ \bibinfo {pages} {030336}
  (\bibinfo {year} {2023})}\BibitemShut {NoStop}%
\bibitem [{\citenamefont {Khaneja}\ \emph {et~al.}(2005)\citenamefont
  {Khaneja}, \citenamefont {Reiss}, \citenamefont {Kehlet}, \citenamefont
  {Schulte-Herbr{\"u}ggen},\ and\ \citenamefont {Glaser}}]{Khaneja2005}%
  \BibitemOpen
  \bibfield  {author} {\bibinfo {author} {\bibfnamefont {N.}~\bibnamefont
  {Khaneja}}, \bibinfo {author} {\bibfnamefont {T.}~\bibnamefont {Reiss}},
  \bibinfo {author} {\bibfnamefont {C.}~\bibnamefont {Kehlet}}, \bibinfo
  {author} {\bibfnamefont {T.}~\bibnamefont {Schulte-Herbr{\"u}ggen}}, \ and\
  \bibinfo {author} {\bibfnamefont {S.~J.}\ \bibnamefont {Glaser}},\ }\bibfield
   {title} {\enquote {\bibinfo {title} {Optimal control of coupled spin
  dynamics: design of nmr pulse sequences by gradient ascent algorithms},}\
  }\href {https://doi.org/10.1016/j.jmr.2004.11.004} {\bibfield  {journal}
  {\bibinfo  {journal} {J. Magn. Reson.}\ }\textbf {\bibinfo {volume} {172}},\
  \bibinfo {pages} {296} (\bibinfo {year} {2005})}\BibitemShut {NoStop}%
\bibitem [{\citenamefont {Kubica}\ \emph {et~al.}(2023)\citenamefont {Kubica},
  \citenamefont {Haim}, \citenamefont {Vaknin}, \citenamefont {Levine},
  \citenamefont {Brand\~ao},\ and\ \citenamefont {Retzker}}]{Kubica2023}%
  \BibitemOpen
  \bibfield  {author} {\bibinfo {author} {\bibfnamefont {A.}~\bibnamefont
  {Kubica}}, \bibinfo {author} {\bibfnamefont {A.}~\bibnamefont {Haim}},
  \bibinfo {author} {\bibfnamefont {Y.}~\bibnamefont {Vaknin}}, \bibinfo
  {author} {\bibfnamefont {H.}~\bibnamefont {Levine}}, \bibinfo {author}
  {\bibfnamefont {F.}~\bibnamefont {Brand\~ao}}, \ and\ \bibinfo {author}
  {\bibfnamefont {A.}~\bibnamefont {Retzker}},\ }\bibfield  {title} {\enquote
  {\bibinfo {title} {Erasure qubits: Overcoming the ${T}_{1}$ limit in
  superconducting circuits},}\ }\href {\doibase 10.1103/PhysRevX.13.041022}
  {\bibfield  {journal} {\bibinfo  {journal} {Phys. Rev. X}\ }\textbf {\bibinfo
  {volume} {13}},\ \bibinfo {pages} {041022} (\bibinfo {year}
  {2023})}\BibitemShut {NoStop}%
\bibitem [{\citenamefont {James}\ and\ \citenamefont
  {Jerke}(2007)}]{james2007effective}%
  \BibitemOpen
  \bibfield  {author} {\bibinfo {author} {\bibfnamefont {D.~F.}\ \bibnamefont
  {James}}\ and\ \bibinfo {author} {\bibfnamefont {J.}~\bibnamefont {Jerke}},\
  }\bibfield  {title} {\enquote {\bibinfo {title} {Effective hamiltonian theory
  and its applications in quantum information},}\ }\href {\doibase
  10.1139/p07-060} {\bibfield  {journal} {\bibinfo  {journal} {Can. J. Phys.}\
  }\textbf {\bibinfo {volume} {85}},\ \bibinfo {pages} {625} (\bibinfo {year}
  {2007})}\BibitemShut {NoStop}%
\bibitem [{SI()}]{SI}%
  \BibitemOpen
  \bibfield  {title} {\enquote {\bibinfo {title} {Supplementary material},}\
  }\href@noop {} {\ }\BibitemShut {NoStop}%
\bibitem [{\citenamefont {Magnard}\ \emph {et~al.}(2018)\citenamefont
  {Magnard}, \citenamefont {Kurpiers}, \citenamefont {Royer}, \citenamefont
  {Walter}, \citenamefont {Besse}, \citenamefont {Gasparinetti}, \citenamefont
  {Pechal}, \citenamefont {Heinsoo}, \citenamefont {Storz}, \citenamefont
  {Blais},\ and\ \citenamefont {Wallraff}}]{Magnard2018PRL}%
  \BibitemOpen
  \bibfield  {author} {\bibinfo {author} {\bibfnamefont {P.}~\bibnamefont
  {Magnard}}, \bibinfo {author} {\bibfnamefont {P.}~\bibnamefont {Kurpiers}},
  \bibinfo {author} {\bibfnamefont {B.}~\bibnamefont {Royer}}, \bibinfo
  {author} {\bibfnamefont {T.}~\bibnamefont {Walter}}, \bibinfo {author}
  {\bibfnamefont {J.-C.}\ \bibnamefont {Besse}}, \bibinfo {author}
  {\bibfnamefont {S.}~\bibnamefont {Gasparinetti}}, \bibinfo {author}
  {\bibfnamefont {M.}~\bibnamefont {Pechal}}, \bibinfo {author} {\bibfnamefont
  {J.}~\bibnamefont {Heinsoo}}, \bibinfo {author} {\bibfnamefont
  {S.}~\bibnamefont {Storz}}, \bibinfo {author} {\bibfnamefont
  {A.}~\bibnamefont {Blais}}, \ and\ \bibinfo {author} {\bibfnamefont
  {A.}~\bibnamefont {Wallraff}},\ }\bibfield  {title} {\enquote {\bibinfo
  {title} {Fast and unconditional all-microwave reset of a superconducting
  qubit},}\ }\href {\doibase 10.1103/PhysRevLett.121.060502} {\bibfield
  {journal} {\bibinfo  {journal} {Phys. Rev. Lett.}\ }\textbf {\bibinfo
  {volume} {121}},\ \bibinfo {pages} {060502} (\bibinfo {year}
  {2018})}\BibitemShut {NoStop}%
\bibitem [{\citenamefont {Scigliuzzo}\ \emph {et~al.}(2020)\citenamefont
  {Scigliuzzo}, \citenamefont {Bengtsson}, \citenamefont {Besse}, \citenamefont
  {Wallraff}, \citenamefont {Delsing},\ and\ \citenamefont
  {Gasparinetti}}]{Scigliuzzo2020}%
  \BibitemOpen
  \bibfield  {author} {\bibinfo {author} {\bibfnamefont {M.}~\bibnamefont
  {Scigliuzzo}}, \bibinfo {author} {\bibfnamefont {A.}~\bibnamefont
  {Bengtsson}}, \bibinfo {author} {\bibfnamefont {J.-C.}\ \bibnamefont
  {Besse}}, \bibinfo {author} {\bibfnamefont {A.}~\bibnamefont {Wallraff}},
  \bibinfo {author} {\bibfnamefont {P.}~\bibnamefont {Delsing}}, \ and\
  \bibinfo {author} {\bibfnamefont {S.}~\bibnamefont {Gasparinetti}},\
  }\bibfield  {title} {\enquote {\bibinfo {title} {Primary thermometry of
  propagating microwaves in the quantum regime},}\ }\href {\doibase
  10.1103/PhysRevX.10.041054} {\bibfield  {journal} {\bibinfo  {journal} {Phys.
  Rev. X}\ }\textbf {\bibinfo {volume} {10}},\ \bibinfo {pages} {041054}
  (\bibinfo {year} {2020})}\BibitemShut {NoStop}%
\bibitem [{\citenamefont {Xu}\ \emph {et~al.}(2018)\citenamefont {Xu},
  \citenamefont {Cai}, \citenamefont {Ma}, \citenamefont {Mu}, \citenamefont
  {Hu}, \citenamefont {Chen}, \citenamefont {Wang}, \citenamefont {Song},
  \citenamefont {Xue}, \citenamefont {Yin},\ and\ \citenamefont
  {Sun}}]{Xu2018}%
  \BibitemOpen
  \bibfield  {author} {\bibinfo {author} {\bibfnamefont {Y.}~\bibnamefont
  {Xu}}, \bibinfo {author} {\bibfnamefont {W.}~\bibnamefont {Cai}}, \bibinfo
  {author} {\bibfnamefont {Y.}~\bibnamefont {Ma}}, \bibinfo {author}
  {\bibfnamefont {X.}~\bibnamefont {Mu}}, \bibinfo {author} {\bibfnamefont
  {L.}~\bibnamefont {Hu}}, \bibinfo {author} {\bibfnamefont {T.}~\bibnamefont
  {Chen}}, \bibinfo {author} {\bibfnamefont {H.}~\bibnamefont {Wang}}, \bibinfo
  {author} {\bibfnamefont {Y.~P.}\ \bibnamefont {Song}}, \bibinfo {author}
  {\bibfnamefont {Z.-Y.}\ \bibnamefont {Xue}}, \bibinfo {author} {\bibfnamefont
  {Z.-q.}\ \bibnamefont {Yin}}, \ and\ \bibinfo {author} {\bibfnamefont
  {L.}~\bibnamefont {Sun}},\ }\bibfield  {title} {\enquote {\bibinfo {title}
  {Single-loop realization of arbitrary nonadiabatic holonomic single-qubit
  quantum gates in a superconducting circuit},}\ }\href {\doibase
  10.1103/PhysRevLett.121.110501} {\bibfield  {journal} {\bibinfo  {journal}
  {Phys. Rev. Lett.}\ }\textbf {\bibinfo {volume} {121}},\ \bibinfo {pages}
  {110501} (\bibinfo {year} {2018})}\BibitemShut {NoStop}%
\bibitem [{\citenamefont {Gambetta}\ \emph {et~al.}(2012)\citenamefont
  {Gambetta}, \citenamefont {C\'orcoles}, \citenamefont {Merkel}, \citenamefont
  {Johnson}, \citenamefont {Smolin}, \citenamefont {Chow}, \citenamefont
  {Ryan}, \citenamefont {Rigetti}, \citenamefont {Poletto}, \citenamefont
  {Ohki}, \citenamefont {Ketchen},\ and\ \citenamefont
  {Steffen}}]{Gambetta2012PRL_PTM}%
  \BibitemOpen
  \bibfield  {author} {\bibinfo {author} {\bibfnamefont {J.~M.}\ \bibnamefont
  {Gambetta}}, \bibinfo {author} {\bibfnamefont {A.~D.}\ \bibnamefont
  {C\'orcoles}}, \bibinfo {author} {\bibfnamefont {S.~T.}\ \bibnamefont
  {Merkel}}, \bibinfo {author} {\bibfnamefont {B.~R.}\ \bibnamefont {Johnson}},
  \bibinfo {author} {\bibfnamefont {J.~A.}\ \bibnamefont {Smolin}}, \bibinfo
  {author} {\bibfnamefont {J.~M.}\ \bibnamefont {Chow}}, \bibinfo {author}
  {\bibfnamefont {C.~A.}\ \bibnamefont {Ryan}}, \bibinfo {author}
  {\bibfnamefont {C.}~\bibnamefont {Rigetti}}, \bibinfo {author} {\bibfnamefont
  {S.}~\bibnamefont {Poletto}}, \bibinfo {author} {\bibfnamefont {T.~A.}\
  \bibnamefont {Ohki}}, \bibinfo {author} {\bibfnamefont {M.~B.}\ \bibnamefont
  {Ketchen}}, \ and\ \bibinfo {author} {\bibfnamefont {M.}~\bibnamefont
  {Steffen}},\ }\bibfield  {title} {\enquote {\bibinfo {title}
  {Characterization of addressability by simultaneous randomized
  benchmarking},}\ }\href {\doibase 10.1103/PhysRevLett.109.240504} {\bibfield
  {journal} {\bibinfo  {journal} {Phys. Rev. Lett.}\ }\textbf {\bibinfo
  {volume} {109}},\ \bibinfo {pages} {240504} (\bibinfo {year}
  {2012})}\BibitemShut {NoStop}%
\bibitem [{\citenamefont {Tsunoda}\ \emph {et~al.}(2023)\citenamefont
  {Tsunoda}, \citenamefont {Teoh}, \citenamefont {Kalfus}, \citenamefont
  {de~Graaf}, \citenamefont {Chapman}, \citenamefont {Curtis}, \citenamefont
  {Thakur}, \citenamefont {Girvin},\ and\ \citenamefont
  {Schoelkopf}}]{Tsunoda2023PRXQ}%
  \BibitemOpen
  \bibfield  {author} {\bibinfo {author} {\bibfnamefont {T.}~\bibnamefont
  {Tsunoda}}, \bibinfo {author} {\bibfnamefont {J.~D.}\ \bibnamefont {Teoh}},
  \bibinfo {author} {\bibfnamefont {W.~D.}\ \bibnamefont {Kalfus}}, \bibinfo
  {author} {\bibfnamefont {S.~J.}\ \bibnamefont {de~Graaf}}, \bibinfo {author}
  {\bibfnamefont {B.~J.}\ \bibnamefont {Chapman}}, \bibinfo {author}
  {\bibfnamefont {J.~C.}\ \bibnamefont {Curtis}}, \bibinfo {author}
  {\bibfnamefont {N.}~\bibnamefont {Thakur}}, \bibinfo {author} {\bibfnamefont
  {S.~M.}\ \bibnamefont {Girvin}}, \ and\ \bibinfo {author} {\bibfnamefont
  {R.~J.}\ \bibnamefont {Schoelkopf}},\ }\bibfield  {title} {\enquote {\bibinfo
  {title} {Error-detectable bosonic entangling gates with a noisy ancilla},}\
  }\href {\doibase 10.1103/PRXQuantum.4.020354} {\bibfield  {journal} {\bibinfo
   {journal} {PRX Quantum}\ }\textbf {\bibinfo {volume} {4}},\ \bibinfo {pages}
  {020354} (\bibinfo {year} {2023})}\BibitemShut {NoStop}%
\bibitem [{\citenamefont {Sun}\ \emph {et~al.}(2014)\citenamefont {Sun},
  \citenamefont {Petrenko}, \citenamefont {Leghtas}, \citenamefont {Vlastakis},
  \citenamefont {Kirchmair}, \citenamefont {Sliwa}, \citenamefont {Narla},
  \citenamefont {Hatridge}, \citenamefont {Shankar}, \citenamefont {Blumoff},
  \citenamefont {Frunzio}, \citenamefont {Mirrahimi}, \citenamefont {Devoret},\
  and\ \citenamefont {Schoelkopf}}]{SunNature}%
  \BibitemOpen
  \bibfield  {author} {\bibinfo {author} {\bibfnamefont {L.}~\bibnamefont
  {Sun}}, \bibinfo {author} {\bibfnamefont {A.}~\bibnamefont {Petrenko}},
  \bibinfo {author} {\bibfnamefont {Z.}~\bibnamefont {Leghtas}}, \bibinfo
  {author} {\bibfnamefont {B.}~\bibnamefont {Vlastakis}}, \bibinfo {author}
  {\bibfnamefont {G.}~\bibnamefont {Kirchmair}}, \bibinfo {author}
  {\bibfnamefont {K.~M.}\ \bibnamefont {Sliwa}}, \bibinfo {author}
  {\bibfnamefont {A.}~\bibnamefont {Narla}}, \bibinfo {author} {\bibfnamefont
  {M.}~\bibnamefont {Hatridge}}, \bibinfo {author} {\bibfnamefont
  {S.}~\bibnamefont {Shankar}}, \bibinfo {author} {\bibfnamefont
  {J.}~\bibnamefont {Blumoff}}, \bibinfo {author} {\bibfnamefont
  {L.}~\bibnamefont {Frunzio}}, \bibinfo {author} {\bibfnamefont
  {M.}~\bibnamefont {Mirrahimi}}, \bibinfo {author} {\bibfnamefont {M.~H.}\
  \bibnamefont {Devoret}}, \ and\ \bibinfo {author} {\bibfnamefont {R.~J.}\
  \bibnamefont {Schoelkopf}},\ }\bibfield  {title} {\enquote {\bibinfo {title}
  {Tracking photon jumps with repeated quantum non-demolition parity
  measurements},}\ }\href {\doibase 10.1038/nature13436} {\bibfield  {journal}
  {\bibinfo  {journal} {Nature}\ }\textbf {\bibinfo {volume} {511}},\ \bibinfo
  {pages} {444} (\bibinfo {year} {2014})}\BibitemShut {NoStop}%
\end{thebibliography}
\end{document}